\documentclass[%
 aip,
 amsmath,amssymb,
 reprint,%
]{revtex4-1}

\usepackage{graphicx}
\usepackage{dcolumn}
\usepackage{bm}
\usepackage{booktabs}
\usepackage{makecell}

\usepackage[utf8]{inputenc}
\usepackage[T1]{fontenc}
\usepackage{mathptmx}
\usepackage{etoolbox}
\usepackage{color}
\newcommand{\Vm}{V_\text{m}}
\newcommand{\mV}{\text{mV}}

\makeatletter
\def\@email#1#2{%
 \endgroup
 \patchcmd{\titleblock@produce}
  {\frontmatter@RRAPformat}
  {\frontmatter@RRAPformat{\produce@RRAP{*#1\href{mailto:#2}{#2}}}\frontmatter@RRAPformat}
  {}{}
}%
\makeatother
\begin{document}

\preprint{AIP/123-QED}

 	\title{
  Annihilation dynamics during
spiral defect chaos revealed by particle models}
	\author{Timothy J Tyree}
	\affiliation{Department of Physics, University of California, San Diego, CA}
	\author{Patrick Murphy}
	\affiliation{Department of Mathematics and Statistics, San Jose State University, San Jose, CA}
	\author{Wouter-Jan Rappel}
	\affiliation{Department of Physics, University of California, San Diego, CA}

	\date{\today}
	
	\begin{abstract}
		Pair-annihilation events are ubiquitous in a variety of spatially extended systems and are often studied using computationally expensive simulations.  Here we develop an  approach in which
		we simulate the pair-annihilation of spiral wave tips in cardiac models using
		a computationally efficient particle model.
		Spiral wave tips
		are represented as particles with dynamics governed by
		diffusive behavior and short-ranged attraction.
		The parameters for diffusion and attraction are obtained by comparing
		particle motion to the trajectories of spiral wave tips
		in cardiac models during spiral defect chaos.
		The particle model 
		reproduces the annihilation rates of the cardiac models and
		can determine the statistics of spiral wave dynamics, including its
		mean termination time. We show that increasing the attraction coefficient
		sharply decreases the mean termination time, making it a possible
		target for pharmaceutical intervention.
	\end{abstract}
	
	\maketitle
	
	\begin{quotation}
		Many physical systems exhibit annihilation events during which
		pairs of objects collide and are removed from the system. 
		These events occur in a number of soft-matter and active-matter systems that exhibit spatiotemporal patterning.  For example, topological defects in nematic liquid crystals
		can develop motile topological defects that annihilate when they meet ~\cite{doostmohammadi2018active,liu2007dynamics}.
		Pair-wise annihilation of defects or singularities also plays a role in a number 
		of biological systems. 
		In bacterial biofilms, for instance, 
		imperfect cell alignment results in   
		point-like defects that carry half-integer topological charge and can 
		annihilate in pairs. 
		These topological defects 
		explain the formation of layers and 
		have been proposed as a model for the buckling of biofilms in colonies of nematically ordered cells~\cite{boyer2011buckling,copenhagen2021topological}.
	\end{quotation}
	
	\section{Introduction}
	
	In this study, we focus on the
	pair-wise annihilation that occurs during spiral defect chaos in excitable 
	systems. During spiral defect chaos, spiral waves
	continuously break down to form new
	ones and are removed through collisions with other spiral waves.
	While spiral defect chaos is present in a variety of  chemical and biological pattern-forming systems \cite{coullet1989defect, rehberg1989traveling, ecke1995excitation, hildebrand1995statistics, ouyang1996transition, Egoetal00, DanBod02,varela2005transitions,beta2006defect,qu2006critical,hofer1995dictyostelium,epstein1996nonlinear}, perhaps its most studied example can be found in models of 
	cardiac tissue 
	\cite{karma1994electrical,ten2004model, clayton2011models, rappel2022thephysics}.
	These models naturally exhibit  
	spiral waves and the tips of these spiral waves undergo 
	stochastic annihilation events~\cite{lilienkamp2018terminal,vidmar2019extinction,lilienkamp2020terminating}. 
	Importantly, these annihilations underlie cardiac fibrillation, characterized by unorganized electrical wave propagation in the heart \cite{rappel2022thephysics}. 
	Fibrillation in the ventricles is  
	lethal \cite{jalife2000ventricular} while 
	atrial fibrillation, the most common cardiac arrhythmia in the world with approximately 30 million patients in 2010, is
	associated with increased morbidity and mortality \cite{Nacetal09,miyasaka2007mortality,chugh2014worldwide,lane2017temporal}.
	
	Previous studies have described the statistics of spiral wave tips in 
	spatially extended
	systems in the form of a stochastic birth-death process 
	\cite{gil1990statistical, DanBod02, beta2006defect}.  
	More recently, this approach was applied to cardiac models, where  
	the creation and annihilation rates of spiral tips were determined for various domain sizes \cite{vidmar2019extinction,rappel2022stochastic}.
	Since 
	$N=0$ is an absorbing state, the dynamical state will eventually terminate.   
	Using these rates,  
	the mean termination time,  $\tau$, was computed and was shown to be exponentially distributed, consistent with experiments and clinical data  \cite{dharmaprani2019renewal,vidmar2019extinction,rappel2022stochastic}. 
	This termination time
	is a quantity of interest in the context of 
	cardiac dynamics as termination indicates the heart 
	has transitioned into normal sinus rhythm.  Thus, minimizing $\tau$ is of critical importance for managing cardiac fibrillation.
	
	Previous work has shown $\tau$ depends
	on the tissue size, $A$, 
	and reducing $\tau$ can be achieved by decreasing $A$
	\cite{qu2006critical,vidmar2019extinction}. 
	Unfortunately, decreasing the effective size of cardiac tissue is not practical and determining the dependence of $\tau$ on other physiological properties is therefore desirable, especially if these properties can be drug-targeted.
	In this study, we propose targeting the attraction coefficient, $a$, which controls the strength of attraction between spiral wave tips.  
	
	We first simulate two spatially extended cardiac models and show that the spiral tips that annihilate in these models display an apparent attractive interaction at short distances and diffusive Brownian behavior at large distances. 
	We then formulate a stochastic particle model in which 
	tips are represented as particles and show that it 
	can capture the attractive and diffusive properties of the tips
	in the cardiac models. 
	Furthermore, we show that this particle model
	generates tip dynamics that reproduce 
	both the annihilation rates as a function 
	of the density of tips and
	the distribution of termination times for the two cardiac models. 
	Finally, we show that increasing the attraction coefficient significantly decreases the mean termination time of spiral defect chaos. 
	
	\section{Cardiac Models}
	
	To determine the dynamics of spiral wave tips in the cardiac models, we
	integrated the mono-domain equations, which 
	describe the time evolution of the membrane potential, $u$, by
	the excitable reaction-diffusion equation
	\begin{equation}\label{eqn:eom_volt}
		\partial_tu= D_{u}\nabla^2 u -I_\text{ion}/C_\text{m}
	\end{equation}
	\noindent where $I_\text{ion}$ represents the transmembrane currents, $C_\text{m}=1\mu\text{F}/\text{cm}^2$ is the transmembrane capacitance per unit area, and $D_{u}$ is the 
	diffusion coefficient
	\cite{rappel2022thephysics}. 
	To stress the generality of our approach, we used two 
	commonly-employed models for cardiac tissue: the Fenton-Karma (FK) model~\cite{FenKar98} and the Luo-Rudy (LR) model~\cite{LuoRud91}.
	The precise formulation of $I_\text{ion}$ for these models is provided in Appendix A, along with model parameters. 
	
	We integrated Eq.~\ref{eqn:eom_volt}  on a square computational domain of size $A$ and enforced
	periodic boundary conditions.
	These periodic boundary conditions resulted only in pair-creation and pair-annihilation of spiral tips 
	due to conservation of vorticity in $u$ and, thus, in even numbers of 
	tips.
	We used a spatial discretization of $\Delta x=0.025$ cm and a temporal discretization of $\Delta t=0.025$ms and used a
	body-centered forward-time explicit Euler method with the Laplacian operator discretized using a five-point stencil on the square lattice. 
	
	We  chose parameter values for which both models exhibit spiral defect chaos. In other words, a single spiral wave is unstable and will break up into multiple spiral waves.
	Observations of spiral tip motion began 100 ms after the start of the simulation at time $t=0$ so as to allow periodic boundaries enough time to become smooth.
	The locations of spiral tips were determined from the intersection points of two level sets of constant voltage ($u=0.4$ for the FK model and $u=-30$ mV for the LR model).  One level set was the equipotential line at the constant threshold voltage.  The other level set was the equipotential line at the constant threshold voltage 4 ms later in time.  Linear segments of level sets were determined by linear interpolation along the edges that connected nodes using the marching squares method described by Lewiner et al. \cite{lewiner2003efficient} modified to support the periodic boundary conditions.  Intersection points were determined by solving the linear system of equations that describes two intersecting linear segments and were recorded as spiral tip locations.
	For the chosen parameter values, the width of a planar wave was 0.41 cm for the FK model and 0.15 cm for the LR model, computed at the same thresholds used for the detection of spiral tip locations.
	
	\section{Cardiac Model Results}
	
\begin{figure}[h!]
		\begin{center}
		\includegraphics[width=\columnwidth]{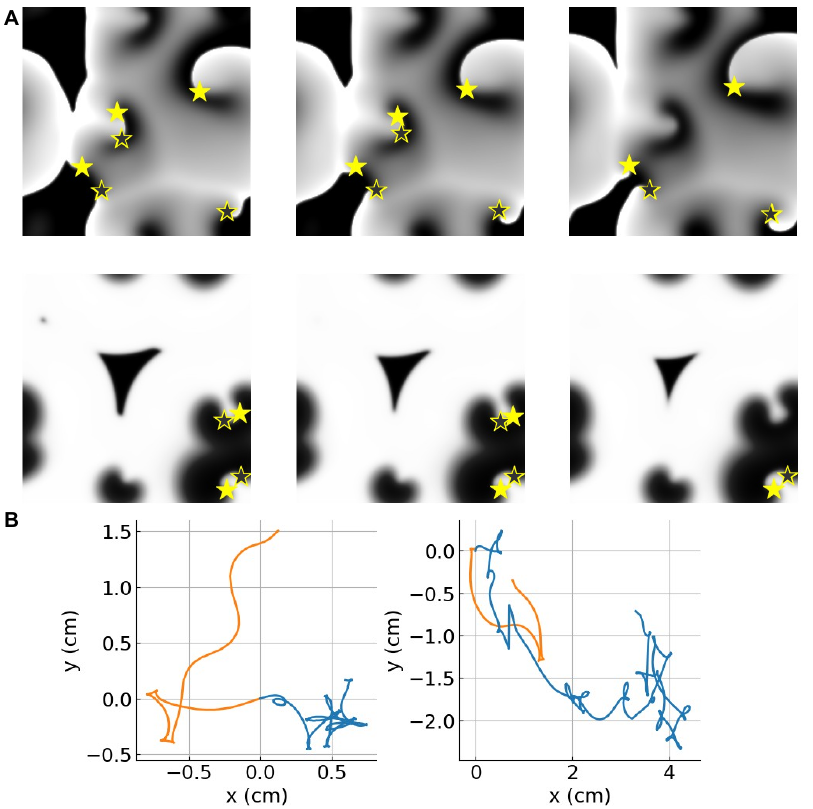}
			\caption{ \textbf{A} 
				Grayscale snapshots of $u$ showing spiral defect chaos in (top) the LR model and (bottom) the FK model with  $A=25\text{ cm}^2$.  Indicated are the tips of  clockwise (black stars) and counterclockwise (yellow stars)  rotating spiral waves.  Snapshots were taken at (left) $t’=8$ ms, (middle) $t’=4$ ms, and
				(right) $t’=0$ ms before an annihilation event.   
				Annihilation can be explained by
				a wave-block resulting from a depolarized area acting as
				a wall to spiral tip motion.
				\textbf{B} Single tip trajectories are shown from the FK model and  the LR model (orange). The trajectories in the LR model show fewer pivots, supporting our result that the LR model has a larger diffusion coefficient than the FK model 
				(see also Table~\ref{tab:properties}).
				}
			\label{fig:dynamics}
		\end{center}
	\end{figure}

	Snapshots of typical simulations displaying a pair-wise
	annihilation event are shown 
	in Fig.~\ref{fig:dynamics}A
	for the LR model (upper row) and the FK model (lower row). 
	In these panels,  $u$ is visualized using a grayscale and clockwise and counterclockwise tips are indicated by 
	black and yellow symbols, respectively.
	Examples of single spiral tip trajectories are 
	shown in 
	Fig.~\ref{fig:dynamics}B, with the FK trajectories plotted in 
	blue and the LR trajectories in orange.
 Although spiral tips can be both created and annihilated, we focus here on the 
 behavior of annihilating spiral wave tips.
	In our simulations, only pairs of counter-rotating spiral waves that 
	are connected by an activation front can annihilate.
	Annihilation occurs when a depolarized region acts as a wave block, causing the activation front to shrink in length before spontaneously disappearing.

For both models, we examined whether 
tips that came within a certain distance were able to escape and 
move apart or always annihilated. Specifically, we determined the number of pairs that became separated by less than 0.1 cm and were then able to 
 move further apart than 0.15 cm. These pairs always consisted 
 of spiral waves of opposite chirality: one was rotating clockwise while the 
 other was rotating counterclockwise \cite{ermakova1989interaction,kalita2022interaction}. Examining 355 pairs in the FK model and 99 pairs in the LR revealed that these pairs always annihilated.  
 In the remainder of this study, we will strictly focus on annihilation events and
 all subsequent results are therefore conditioned on these dynamics.

\begin{figure}
		\begin{center}
		\includegraphics[width=\columnwidth]{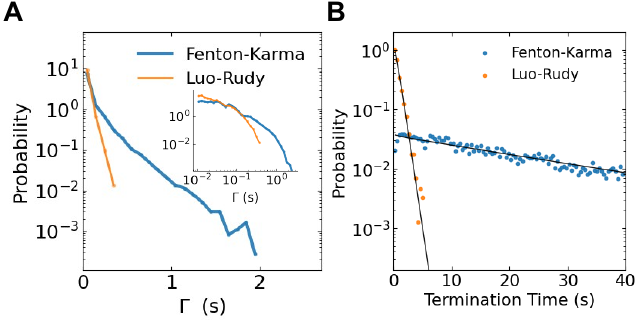}
			\caption{ 
				\textbf{A} Probability density of  lifetime $\Gamma$ of annihilating 
				pairs in the two cardiac models. Inset is the probability density of $\Gamma$ to visualize the short-lived tips, revealing that the  abundance of short-lived spiral tips ($<$ 15ms) was somewhat greater for the LR model relative to the FK model.
				This log-log histogram has 10 bins per decade.
				\textbf{B} Probability density of the termination times of 
				simulations of the two cardiac models, along with an exponential fit (solid line).  
			}
			\label{fig:extinction}
		\end{center}
	\end{figure}

	To quantify the dynamics of annihilating spiral tips,
	we first computed the lifetime of  pairs that annihilate, $\Gamma$.
	For this, pair-annihilation events were determined by ordinary nearest-neighbor particle tracking subject to periodic boundary conditions. 
	We computed $\Gamma$ as the time of annihilation minus the time of creation of the younger of the pair.
	The distribution of these lifetimes was approximately exponential (Fig.~\ref{fig:extinction}A) with the FK model producing significantly longer-lived pairs on average (Table~\ref{tab:properties}).

	The distribution of termination times of spiral defect chaos, 
	i.e. the time until all spiral tips have been removed, was
	also found 
	to be fitted well with an exponential  
	(Fig.~\ref{fig:extinction}B, see also \cite{vidmar2019extinction}).
	This can be understood by considering not only pair 
	annihilation but also pair creation, which may occur when a wave back meets a wave front. 
	As a result of this continuous creation and annihilation 
	process, the number of tips, $N$, reaches a long-lived 
	metastable state \cite{dykman1994large,assaf2010extinction}. 
	The distribution associated with this metastable state
	is called the  quasi-stationary distribution and can 
	be calculated using the 
	master equation for the probability $P(N,t)$ to have $N=0,2,4,8,...$ 
	tips at time $t$ \cite{vidmar2019extinction}:
	\begin{eqnarray}
		\frac{dP (N,t)}{dt} &=& W_{-2}(N+2)P(N+2,t)-W_{-2}(N)P(N,t)\label{eqn:master} \\
		& & \;+\; W_{+2}(N-2)P(N-2,t)-W_{+2}(N)P(N,t)\,, \nonumber
	\end{eqnarray}
	where $W_{\pm 2}$ are the annihilation and creation rates.
	If the number of spiral tips reaches zero, the 
	spiral defect chaos has terminated. In other words, $N=0$ in the above
	equation is an absorbing boundary, and  
	$W_{+2}(0)=0$. Through
	repeated simulations of  termination events,  we 
	determined the distribution of termination times, which is 
	shown for both models in Fig.~\ref{fig:extinction}B.  
	
\begin{figure}[h]
		\begin{center}
		\includegraphics[width=\columnwidth]{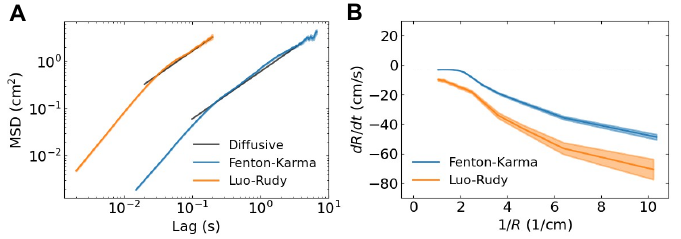}
			\caption{ 
				\textbf{A} MSD of spiral tips \textit{versus} temporal lag.  Black lines indicate Brownian motion.  
				\textbf{B} Mean radial velocity \textit{versus} inverse distance between annihilating tips.
				Shaded bands represent 95\% confidence intervals. Dashed line represents $dR/dt=0$. 
			}\label{fig:MSD_dRdt}
		\end{center}
	\end{figure}

\begin{figure}
		\begin{center}
		\includegraphics[width=\columnwidth]{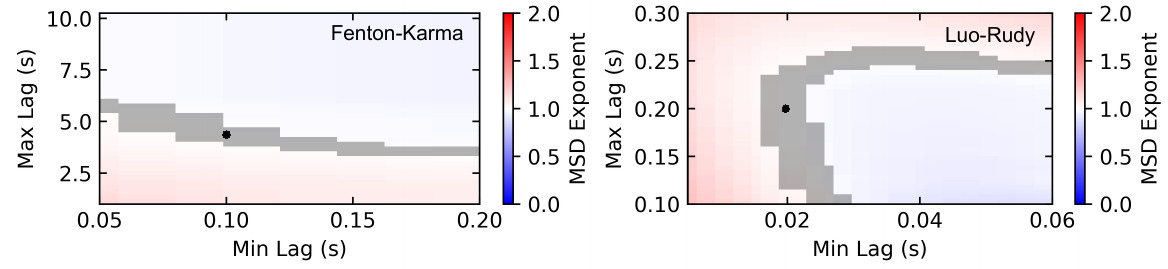}
			\caption{
				Exponents of power laws fit to MSD plotted as a function of lag window for  the FK model (left) and  the LR model (right). The resulting exponents are visualized using a color scale and 95\% confidence intervals were determined by ordinary least squares. Gray shaded regions indicate where the 95\% confidence interval contained the slope of 1, which corresponds to Brownian motion.  The remaining regions exhibited a statistically significant difference from Brownian behavior ($p<0.05$).
				Indicated by the black dots are the fits reported in the text.
			}
			\label{fig:msd_exp}
		\end{center}
	\end{figure}

	By averaging
	6,043 tip trajectories at a domain size of $A=25\text{ cm}^2$, we next  
	computed the mean squared displacement (MSD) as a function of time lag \cite{qian1991single} (Fig.~\ref{fig:MSD_dRdt}A).  
	For both cardiac models, 
	the MSD was not significantly different from linear for long timescales 
	with exponent values of $1.002\pm0.012$ (FK model for time lags from 100ms to 4000ms)
	and $0.996\pm0.019$ (LR model for time lags from 20 ms to 200 ms). 
	Multiple other fits using different time lag windows supported this finding (Fig.~\ref{fig:msd_exp}).

	Thus, 
	in both cardiac models, the spiral wave tips can be described as effectively undergoing 
	Brownian motion for long timescales.  At short timescales, however, tips did not exhibit diffusive behavior.  
	To determine the behavior of spiral tips at these short timescales, 
	we analyzed the movement of annihilating spiral waves in the simulations. 
	This analysis revealed that annihilation occurs when the activation front connecting the tip pair is blocked by a nearby polarized region. 
	The block results in a rapid shrinking of the activation front and the removal of the pair (see Movie S1). 
	Thus, at short timescales, the annihilating pair of tips appear to attract, which becomes apparent from Fig.~\ref{fig:MSD_dRdt}B where 
	we plot the ensemble-averaged radial velocity $dR/dt$ as a function of the 
	distance between the tips, $R$.
	This velocity is
	not constant, but instead
	becomes more negative when $1/R$ increases (and $R$ decreases). 
	In other words, an apparent attractive force induced the annihilation of counter-rotating pairs of spiral waves. 
 We should stress that this force 
 is only apparent and does not arise from a physical attraction between the 
 different spiral waves. 
	
\begin{figure}[h]
		\begin{center}
    \includegraphics[width=0.5\columnwidth]{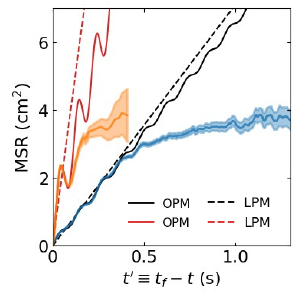}
        \caption{ 
				\textbf MSR between annihilating tips \textit{versus} time
				until annihilation from simulations of the FK and
				LR models with shaded regions
				corresponding to 95\% confidence intervals. 
				Also shown are the fits of MSR from the OPM (solid lines) and the LPM (dashed lines). 
			}\label{fig:MSR}
		\end{center}
	\end{figure}

	We next sought to quantify this attractive force. 
	Since we were interested in the attractive force during annihilation, we only focused on 
	pair-wise annihilation events in our simulations, 
	ignoring creation events. We identified 51,452 annihilating pairs and,
	for a given pair annihilating
	at time $t_f$, we determined $R$
	\textit{versus} time until annihilation, $t'\equiv t_f - t \ge 0$. 
	From this, we computed the mean squared range (MSR) 
	by ensemble-averaging $R^2$ conditioned on a given $t'$.
	The results of this ensemble-averaging are shown in Fig.~\ref{fig:MSR} as solid lines, together with 
	the 95\% confidence intervals as shaded areas. 
	On short timescales, the MSR demonstrates oscillations, 
	illustrating that the apparent attractive force has an oscillatory component.
	As will be discussed, this oscillatory component is simply the manifestation of the rotation of the spirals.
	
	\section{Stochastic Particle Models}
	
	\subsection{Oscillatory Particle Model}
	
	To gain further insight into the annihilation dynamics of the tips using the spatially extended cardiac models 
	is computationally expensive, especially for large domain sizes. 
	Therefore, 
	we developed a model in which the spiral wave tips are represented by  
	moving particles subject to an oscillatory short-range attractive force and 
	Brownian motion with diffusion coefficient, $D$.
	Such a significant simplification of a cardiac model was earlier used to 
	describe the chaotic tip trajectories of a single spiral wave in the 
	presence of heterogeneities
	\cite{lombardo2019chaotic}.
	
	In this oscillatory particle model
	(OPM), we modeled the attractive force between two annihilating tips as
	inversely proportional to the distance between them. 
	This attractive force was assumed to consist of a constant term, parameterized 
	by the coefficient $a_0$,  together with an oscillatory term,
	parameterized by $a_1$.
	This oscillatory term arises from the fact that each  spiral wave rotates 
and that the attractive force depends on the relative location of the two activation fronts
	\cite{ermakova1989interaction, kalita2022interaction}. 
 This relative location is determined by the angle of the activation front (AAF), $\theta$, with an arbitrary, fixed axis of each spiral wave. 
	For example, one could choose to measure $\theta$ relative to the positive x-axis.
	Measuring this AAF, however, is challenging since it 
	requires determining the exact location of both the tip and the activation front and is thus prone to large discretization errors. For this reason, we will not explicitly
 quantify the AAF in this study. 
 
 Based on observations 
	of the cardiac model simulations, we will take  the frequency $\omega$ and period $T_{OPM}$ of the annihilating pair to be identical. 
	Furthermore, as we pointed out before, the spiral waves of an annihilating pair have opposite chirality 
	so that AAF of the first particle progresses as $\theta_1(t)=\omega t + \theta_1$ and 
	that of the second one as $\theta_2(t)=-\omega t + \theta_2 $. 
	Then, the  difference between the two AAFs can be written as $\Delta \theta/2= \omega t + \theta_f$, where $\theta_f=(\theta_1-\theta_2)/2$.
 This difference between the two AAFs  
	is maximal and minimal once during a complete rotation of the two tips. Based 
 on the oscillatory nature of the MSR curves in Fig. \ref{fig:MSR},
  we will assume 
 that the attractive force depends on this difference.
	Taken together, the deterministic part of the OPM for one of the 
	particles, located  at $(X_1,Y_1)$,
	is written as 
	\begin{eqnarray}
		\frac{dX_1}{dt} =  \frac{X_2-X_1}{R^2}\Big(a_0+a_1\cos\big(\omega t + \theta_f\big)\Big) \\
		\frac{dY_1}{dt} =  \frac{Y_2-Y_1}{R^2}\Big(a_0+a_1\cos\big(\omega t + \theta_f\big)\Big)
	\end{eqnarray}
	where $R=\sqrt{(X_2-X_1)^2+(Y_2-Y_1)^2}$ is the distance between 
	this particle and the other particle, located at  $(X_2,Y_2)$. 
	The deterministic OPM equations for this other particle are then given by
	$\frac{dX_2}{dt} = -\frac{dX_1}{dt}$, $\frac{dY_2}{dt} = - \frac{dY_1}{dt}$.
	
	To capture the diffusive motion of the spiral tips, we included an independent Brownian motion term to the deterministic equations for each particle.    Then, 
	we find that the  distance between annihilating tips 
	can be modeled
	by the following Langevin equation:
	\begin{equation}\label{eqn:osc_eom}
		dR(t)=-\frac{2}{R(t)}\Big(a_0+a_1\cos\big(\omega t + \theta_f\big)\Big)dt + \sqrt{8D} dW_{t}\,,
	\end{equation}
	\noindent where $W_{t}$ is a Wiener process at time $t$ and 
	$D$ is estimated from the least-square slope of MSR at large ranges ($\text{MSR}>3\text{ cm}^2$).
	Note that the factor of two arises from the pair-wise interaction. Furthermore, since the variance of the two two-dimensional tips add, the Wiener process is 
	multiplied by $\sqrt{8D}$ instead of $\sqrt{2D}$. 
	
	Using the Langevin equation, it is straightforward to derive an
	expression for the MSR, $<R^2>$. This MSR has both a deterministic and 
	a stochastic term:
	
 \begin{eqnarray}
	    \label{eqn:msrosc}
		\text{MSR}_\text{OPM}(t')=4\Big(a_0 t'+\frac{a_1}{\omega} 
		\big(\sin(\omega t' + \theta_f) - \nonumber \\
		\sin(\theta_f)\big)+2Dt'\Big). 
  \end{eqnarray}
	
	\noindent
	This result was verified by comparing this expression to the average MSR of
	10,000 statistically independent explicit simulations of Eq.~\ref{eqn:osc_eom}.
	These simulations can be carried out by either starting with a large value  of $R$ and progressing until $R$ falls below a small threshold value or by 
	starting with a small value for $R$ and integrating backwards in time.
	
	The next step was to fit Eq.~\ref{eqn:msrosc} to the MSR curves obtained from the cardiac models 
	using simulated annealing on the last 300 ms (FK) and 100 ms (LR) before annihilation.
	The fits to both cardiac models for these time intervals are excellent (mean percent error (MPE) $<$4\%), as can be seen 
	Fig.~\ref{fig:MSR}.  The fitted parameters, $a_0,\,a_1,\,\theta_f,\text{ and }T_{OPM}$ are reported in Table~\ref{tab:properties}, together with the aforementioned estimates for $D$.
	In Table~\ref{tab:properties} 
	we also report the period of the spiral waves 
	in the cardiac models, $T$, determined by computing the mean number of rotations per lifetime.  
	A comparison between $T_{OPM}$ and $T$ reveals that these periods match 
	perfectly, indicating that the oscillatory component in the MSR is due to the rotational nature of the spiral wave.  
	
	\subsection{Linear Particle Model}
	
	The OPM can, in principle, be used to compute annihilation rates, which 
	can then be compared to results from the cardiac models.  
	However, computing these rates with the OPM
	is challenging because annihilation only occurs when the final difference in AAF between the two particles takes a specific value. 
	Thus, it requires tediously tracking the AAF for 
	each particle and enforcing the final condition $\theta(t_f)=\theta_f$
	at the time of annihilation. 
	To circumvent this problem, we simplified the OPM to a linear particle model (LPM) by 
	taking the attractive force between particles as linear in $1/R$ and 
	dropping the oscillatory term. 
	In this case, the attractive force is parameterized by a single 
	coefficient $a$ and,   
	modeling the diffusive behavior of spiral tips as before, the LPM reads
	\begin{equation}\label{eqn:sde_linear}
		dR(t)=-\frac{2a}{R}dt + \sqrt{8D} dW_{t}\,,
	\end{equation}\noindent 
	which  results in a MSR given by
	\begin{equation}\label{eqn:msrlin}
		\text{MSR}_\text{LPM}(t')=4(a+2D)t'\,.
	\end{equation}
	
\begin{figure*}
 \includegraphics[width=2\columnwidth]{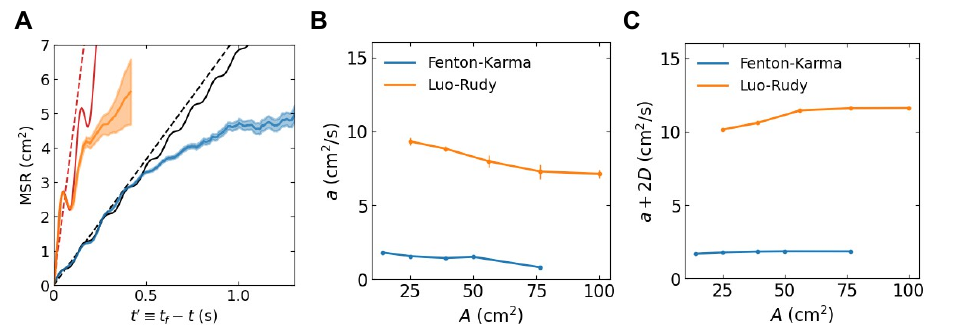}
			\caption{ 
				\textbf{A}   MSR between annihilating tips \textit{versus} time until annihilation from simulations of the
				FK and LR models, using a larger domain 
				size of $A=39.0625\text{ cm}^2$, with shaded regions corresponding to 95\% confidence intervals.
				The solid lines correspond to fits from the OPM while the 
				dashed lines correspond to fits from the LPM. 
				\textbf{B}  Computed attraction coefficient \textit{versus} domain size.
				\textbf{C}  Sum of attractive and diffusive forces \textit{versus} domain size.
				Error bars indicate 95\% confidence.}
			\label{fig:etc}
	\end{figure*}

	To relate $a$ to the parameters of the OPM, 
	we demanded that the MSR averaged over the exponentially distributed lifetimes, $\langle MSR \rangle$,
	be equal for both particle models. For the LPM, we can derive \begin{equation}\label{eqn:R_rms_linear}
		\langle MSR_\text{LPM} \rangle  = \int_0^\infty\Big(\frac{dt'}{\Gamma}e^{-t'/\Gamma}\Big)4(a+2D)t' = 4(a+2D)\Gamma\,.
	\end{equation}
	\noindent  while a similar calculation for the OPM gives 
	\begin{equation}\label{eqn:R_rms_osc}
		\langle MSR_\text{OPM} \rangle =  4\Big(a_0 + a_1 \frac{\cos(\theta_f) - \omega\Gamma\sin(\theta_f)}{1+(\omega\Gamma)^2}+2D\Big)\Gamma\,.
	\end{equation}
	\noindent Setting $\langle MSR_\text{LPM} \rangle =\langle MSR_\text{OPM} \rangle $ results in an analytical expression for $a$ in terms of the 
	OPM parameters that is independent of $D$:
	\begin{equation}\label{eqn:a_estimator}
		a = a_0 + a_1 \frac{\cos(\theta_f) - \omega\Gamma\sin(\theta_f)}{1+(\omega\Gamma)^2}\,.
	\end{equation}
	\noindent The estimates of $a$ evaluated from Eq.~\ref{eqn:a_estimator} are listed in 
	Table~\ref{tab:properties}, and corresponding $\text{MSR}_\text{LPM}$ plots are 
	shown as dashed lines in Fig.~\ref{fig:MSR}. 
	Repeating the analysis for different domain sizes, $A$, revealed that the estimate of $a$ was largely 
	independent of $A$ for both of the cardiac models. 
	An example of   
	the MSR obtained using $A=39.0625\text{ cm}^2$  is shown in Fig.~\ref{fig:etc}A. 
	For this, and all other domain sizes, we were able to fit the MSR using the 
	oscillatory particle model (OPM), as demonstrated by the 
	solid lines in  Fig.~\ref{fig:etc}A, which
	allowed us to compute the value of $a$. 
	The resulting value of $a$ decreased slightly as the domain 
	size increased (Fig.~\ref{fig:etc}B), 
	suggesting that the LPM can be 
	used to simulate spiral tip annihilation at different domain sizes using a fixed set of model parameters determined for a single value of $A$.
	Furthermore, 
	the sum $a+2D$
	also changed little as the domain size increased (Fig.~\ref{fig:etc}C).
	The product of this sum and the mean lifetime 
	is proportional to the mean squared distance (see Eq.~\ref{eqn:R_rms_linear}).
	These results indicate  that  the average distance between annihilating particles is determined by local properties, 
	which are largely insensitive to the domain size.

\begin{table}
\caption{
		Particle properties of spiral tips from the cardiac models including parameter values corresponding to OPM and LPM. 
	}
\begin{ruledtabular}
	\begin{tabular}{lcc}
 \hline
		Symbol & Fenton-Karma &  Luo-Rudy               \\
		\hline
		$\Gamma$ (ms)         & 105.3 $\pm$ 1.7           &  33.4 $\pm$ 0.7         \\
		$D$ (cm$^2$/s)        &  0.115 $\pm$ 0.008        &  0.42 $\pm$ 0.14        \\
		$a_0$ (cm$^2$/s)      &  1.407 $\pm$ 0.016        & 4.2 $\pm$ 0.3           \\
		$a_1$ (cm$^2$/s)      &  1.2822 $\pm$ 0.0005      & 12.180 $\pm$ 0.012      \\
		$\theta_f$ (radians)    & -0.541 $\pm$ 0.004        & -1.165 $\pm$ 0.003      \\
		$T_{OPM}$ (ms)        & 115.94 $\pm$ 0.03         & 97.36 $\pm$ 0.12        \\
		$T$ (ms)              & 115.9 $\pm$ 1.9           & 97.4 $\pm$ 0.8          \\
  	$a$ (cm$^2$/s)        &  1.552 $\pm$ 0.016        &  9.3 $\pm$0.3           \\
		$r$ (cm)              &   0.457$\pm$0.009            &  0.314$\pm$0.003           \\
		$\kappa$ (Hz)         &   15                      &  75                     \\
	\end{tabular}
 \label{tab:properties}
\end{ruledtabular}
\end{table}

	To compute annihilation rates in using particle dynamics instead of 
	cardiac model simulations, we implemented the LPM, using the obtained values of $a$ and $D$, by
	simulating $N$ particles with 
	locations $(X_1(t), Y_1(t), \dots ,X_N(t),Y_N(t))$ on a square domain of size 
	$L^2$ with periodic boundary conditions.
	Each particle $i$ is time stepped with time step $\Delta t=10^{-2}$ms and 
	the $X$ coordinate of particle $i$ obeys
	$$
	X_{i}(t+\Delta t) = X_{i}(t)+
	\sum_{j=1}^N\bigg[a\frac{\big(X_{j}(t)-X_i(t)\big)}{R_{ij}}\bigg] \Delta t +\sqrt{2D\Delta t} Z \,,
	$$
	\noindent where  $D>0$ is the diffusion coefficient, $Z$ is a random value with zero mean and unit variance, and $R_{ij}$ is the distance 
	between particle $i$ and $j$. A similar equation can be written 
	for $Y_i$. 
	For initial conditions, we considered $N$ uniformly distributed particles at two domain sizes ($A=25\text{ cm}^2$ and $A=100\text{ cm}^2$) and updated particle positions every $\Delta t=0.01\text{ ms}$. 
	Pairs of particles were removed from the simulations, and thus 
	annihilated, at rate $\kappa$ whenever they were closer than a 
	reaction range, $r$. As an estimate for  
	this reaction range, we chose the 25$^{th}$ percentile 
	of the distribution of closest distances between non-annihilating tips. We have 
	verified, however, that other choices of $r$ give similar qualitative 
	results.

	\section{Linear Particle Model Results}
	
	\subsection{Annihilation Rates}
	
\begin{figure}[h]
		\begin{center}
    	\includegraphics[width=0.5\columnwidth]{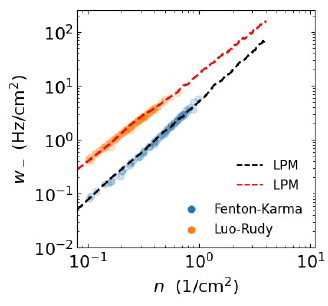}

			\caption{ 
				\textbf  Mean annihilation rate \textit{versus}
				number density for spiral tips from the cardiac
				models (symbols) and their linear particle model fits (dashed lines). 
			}\label{fig:Mean_annihilation}
		\end{center}
	\end{figure}

	We  used the LPM to compute annihilation rates and adjusted the 
	only free parameter ($\kappa$)  to match the annihilation rates found in the cardiac models.   
	This is facilitated by the fact that  
	in our previous work we showed that 
	the latter, computed for different values of $A$, 
	collapsed onto a single curve when rescaled by $A$ \cite{vidmar2019extinction}.  
	Our results show that the annihilation rate can be approximated by a power law, $w_{-} = M_{-} n^{\nu_{-}}$, where $w_- \equiv  W_{-}(N)/A$ is the rescaled annihilation rate and
	$n=N/A$ is the tip density (symbols Fig.~\ref{fig:Mean_annihilation}).
	The fitted LPM annihilation rates are shown as dashed lines in  
	Fig.~\ref{fig:Mean_annihilation} and  resulting parameters are listed in Table~\ref{tab:properties}.
	These fits demonstrate that the LPM can accurately replicate the annihilation rates of the cardiac models (MPE $<$4\%).  
	Importantly, as was the case for $a$, these fits use
	the same parameter values for both system sizes. 
	Also
	note that simulations of the particle model are much more efficient than the cardiac models, especially for 
	large domain sizes. 
	Specifically, the particle model simulations use $\mathcal{O}((L/\Delta x)^2 - N_{avg}^2)$ 
	fewer operations per time step, where $N_{avg}$ is the average number of particles.
	For example, for $A=100$ cm$^2$, the speed-up exceeded $10^4$-fold.
	
	\subsection{Varying Attraction Strength}

\begin{figure}[h!]
		\begin{center}
		\includegraphics[width = \columnwidth]{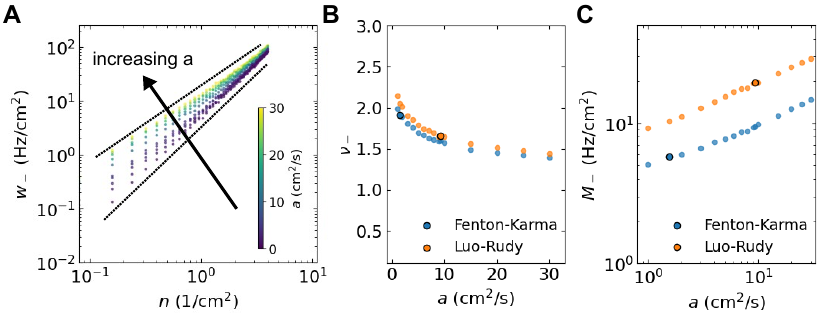}
			\caption{
				\textbf{A} Mean annihilation rate \textit{versus} number density obtained from the LPM using parameters corresponding to the FK model for different values of $a$ (indicated by the inset color bar).
Black lines are guides to the eyes, corresponding to power laws with exponent
4/3 (upper curve) and 2 (lower curve).
				\textbf{B} Power law exponent as a function of $a$ computed using the LPM with parameters corresponding to both the FK and LR model. 
				\textbf{C} Corresponding power law magnitude \textit{versus} $a$.  
				Black circles in B\&C  represent  values of $a$ corresponding to the cardiac models.   Fits considered ordinary least squares over the interval $n\in[0.2,1]$ cm$^{-2}$.}
			\label{fig:a_dependence}
		\end{center}
	\end{figure}

	Although other parameters of the LPM can be varied as well, we focused on  
	the sensitivity of the annihilation rate to the model parameter $a$, 
	which controls the strength of attraction between particles.
	By carrying out additional simulations of the LPM, 
	we determined how the annihilation rate depends on  $a$
	while holding the remaining parameters $D$, $r$, and $\kappa$ fixed.
	This rate was found to increase with increasing values of $a$, which 
	can be understood by realizing that  
	larger attractive forces result in distant particles coming closer together faster.
	Importantly, however, we found that 
	the rate  was  always fitted well by the power law 
	$w_{-} = M_{-} n^{\nu_{-}}$  (Fig.~\ref{fig:a_dependence}A). 
	For both models, we found  
	that for increasing values of $a$ the exponent $\nu_{-}$ became smaller (Fig.~\ref{fig:a_dependence}B) while $M_-$ increased (Fig.~\ref{fig:a_dependence}C).
	
	\subsection{Statistical Properties}
	
\begin{figure}[h]
		\begin{center}
	\includegraphics[width = \columnwidth]{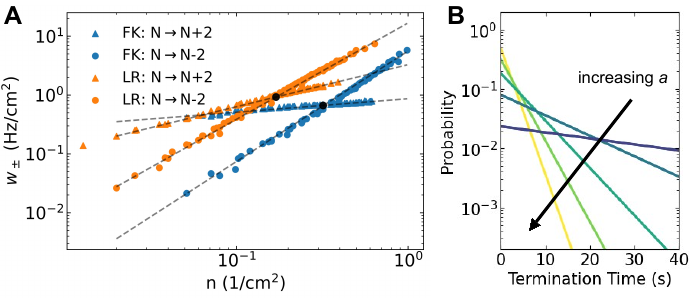}
			\caption{
				\textbf{A} Mean creation rate (triangles) and mean annihilation rate (dots) \textit{versus}
				number density for spiral tips from the
				cardiac models using different domain sizes. Dashed lines correspond to power law fits (Table~\ref{tab:annihil}).
				Black dots correspond to the mean particle density.
				\textbf{B} Probability density of termination times of the LPM for increasing values of $a$ equally spaced from $a=1\text{ cm}^2/\text{s}$ to $a=5\text{ cm}^2/\text{s}$.  Parameter values correspond to the FK model and $A=25\text{ cm}^2$. 
			}\label{fig:statistics}
		\end{center}
	\end{figure}

	Using 
	the fact that the annihilation rate in the LPM can be fitted by a power law,
	we were able to compute statistical properties of spiral defect chaos in the 
	spatially extended cardiac models.  
	For this, we used the creation rate in the cardiac models, which
	were computed  by counting the rate of creation events conditioned on number density. As with the annihilation rates,  
	we found the creation rates to be captured by
	a power law fit, as well: $w_+ = M_+ n^{\nu_+}$ (Fig.~\ref{fig:statistics}A).
	In the remainder, we will keep this 
	creation rate fixed while varying $a$, and thus the annihilation rate, according to Figs.~\ref{fig:a_dependence}B\&C.
	Using the power laws for creation and annihilation rates, we constructed
	a transition matrix from which we computed the distributions of termination times
	\cite{newman2010networks,vidmar2019extinction}.  These
	distributions were found to be exponentially distributed for 
	all values of $a$ (Fig.~\ref{fig:statistics}B), consistent with experimental data  \cite{dharmaprani2019renewal}. 
	Furthermore, 
	the distribution tends to smaller termination time values for increasing $a$. 
	
\begin{table*}
\caption{Power law fits for the annihilation rates and the creation rates. Also included is the mean termination time and average number of particles.
			}\label{tab:annihil}
\begin{ruledtabular}
	\begin{tabular}{lcccccc}
 \hline
Symbol & \multicolumn{3}{c}{Fenton-Karma}& \multicolumn{3}{c}{Luo-Rudy}  \\
\hline
& Cardiac          & LPM                     & LPM             & Cardiac     & LPM         & LPM \\
$A$  (cm$^2$)	 & 25       & 25             & 100     & 25 & 25 & 100  \\
$\nu_-$             & 1.88$\pm$0.03           & 1.871$\pm$0.012 & 1.835$\pm$0.015         &   1.638$\pm$0.017     &  1.614$\pm$0.012 &  1.611$\pm$0.017   \\
$M_-$ (Hz/cm$^2$)   & 5.6$\pm$0.3             & 5.53$\pm$0.16   & 4.75$\pm$0.17           &  16.7$\pm$0.8         & 16.9$\pm$0.7     & 12.6$\pm$0.6       \\
$\nu_+$             & 0.230$\pm$0.010         & 0.230           & 0.230                 &   0.715$\pm$0.010     &  0.715           &  0.715             \\
$M_+$ (Hz/cm$^2$)   & 0.864$\pm$0.002         & 0.864           & 0.864                 &  3.28$\pm$0.10        & 3.28             & 3.28               \\
$\tau$   (s)        &   27.8$\pm$6.5          & 25.9            & 1.68$\times 10^9$     &   0.74$\pm$0.06       &  0.51            &  81.6 \\
$N_{\text{avg}}$    &    8.1$\pm$0.7          & 8.1             &  17.0                 &  5.0$\pm$2.4          & 4.3              &    32.3            \\
\hline
 \end{tabular}
 \end{ruledtabular}
 \end{table*}

	\begin{figure}[h!]
		\begin{center}
	\includegraphics[width=\columnwidth]{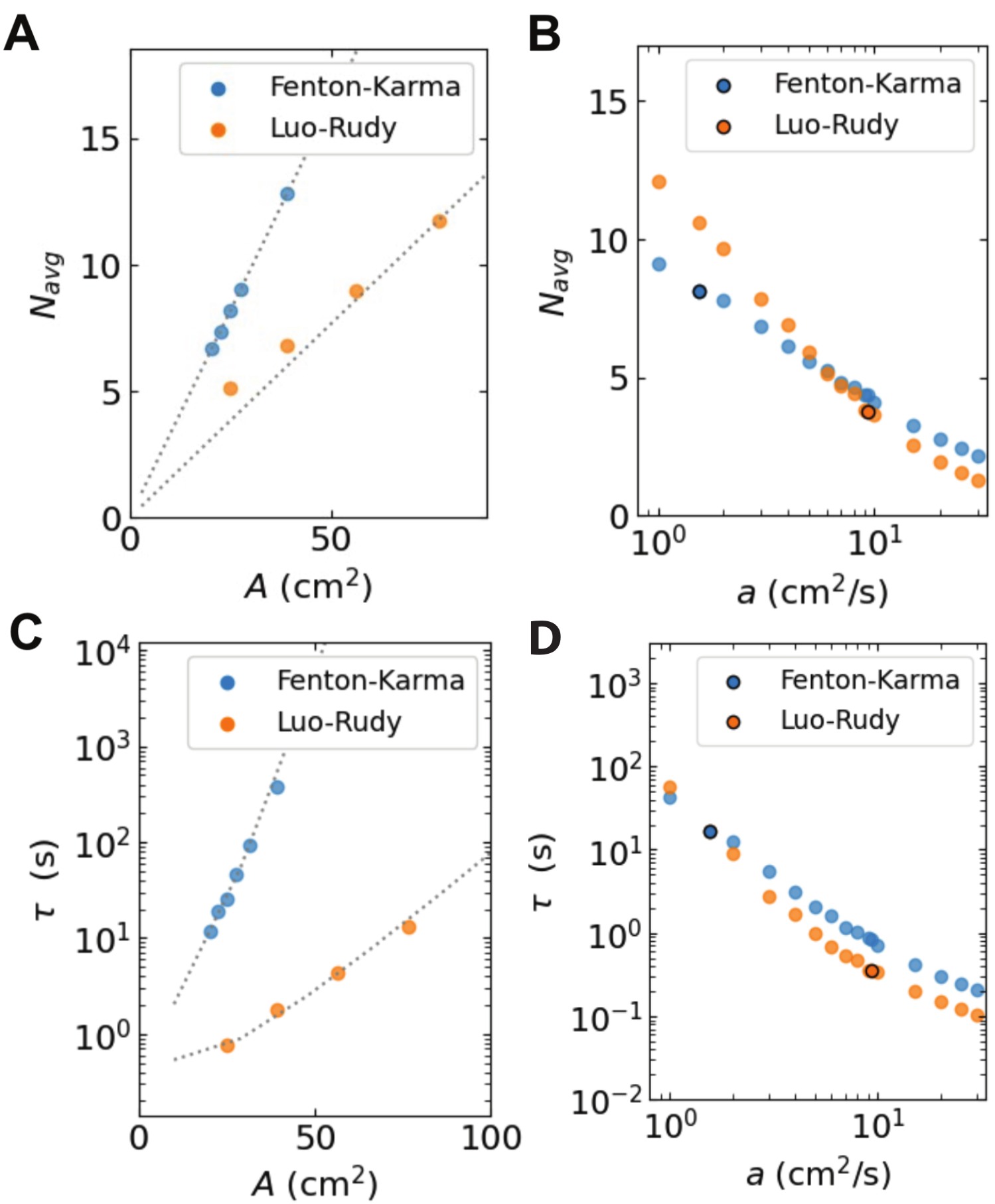}
			\caption{
				\textbf{A} Average tip number as a function of $A$ computed using the 
				cardiac models (symbols), along with the linear prediction of Eq. S\ref{eq:av_number}. 
				\textbf{B} Average tip number as a function of $a$ computed using the 
				LPM with parameter values corresponding to  $A=25\text{ cm}^2$.
				The darkened symbols correspond to the value of $a$ representing the 
				cardiac models.
				\textbf{C} Mean termination time \textit{versus} $A$ computed using Eq. S\ref{scaling} 
				(dashed lines) and separately obtained from the cardiac models (symbols).
				\textbf{D}  Mean termination time as a function of $a$ (using 
				parameter values for  $A=25\text{ cm}^2$).
				Black circles correspond to $a$ obtained by fitting the cardiac models.
			}
			\label{fig:expdeptauonarea}
		\end{center}
	\end{figure}
	
	It is now also possible to obtain approximate expressions for quantities relevant to cardiac models and fibrillation in clinical settings.
	For example, the average number of particles, $N_\text{avg}$  
	can be estimated by solving $w_+(n)=w_-(n)$, resulting in
	\begin{eqnarray}
		N_\text{avg}=A n_\text{avg}=A\left(\frac{M_+}{M_-}\right) ^{\frac{1}{\nu_- - \nu_+}},
		\label{eq:av_number}
	\end{eqnarray}  
	Thus, the mean particle number is predicted to increase linearly with the 
	domain size. 
	Fig.~\ref{fig:expdeptauonarea}A shows the results from Eq.~\ref{eq:av_number}, using 
	parameters corresponding to $A=25$ cm$^2$ for all domain sizes, which is possible 
	since $a$ is insensitive to $A$ (Fig.~\ref{fig:etc}B). 
	The  estimated 
	value using Eq.~\ref{eq:av_number} agrees well with the average number of tips
	from the cardiac models (symbols), especially for larger values of $A$.
 Furthermore, the average number of tips computed using Eq.~\ref{eq:av_number} decreases as a function of $a$ (Fig.~\ref{fig:expdeptauonarea}B), which can 
	be explained by the fact that $\nu_-$ decreases for increasing values of 
	$a$ (Fig.~\ref{fig:a_dependence}B).
	
	In addition, the mean termination time can be computed using 
	an analytic solution  \cite{Gard_book04}.
	Specifically, 
	for paired birth-death processes, the mean termination time 
	$\tau$ is 
	a function of the initial (even) number of spiral tips, $N_0$:
	\begin{eqnarray}
		\tau(N_0)=\sum_{k=1}^{N_0/2}\varphi(2k-2)\sum_{j=k}^{\infty}\frac{1}{\varphi(2j)w_{+}(2j/A)A} 
		\label{eq_mtt_per}
	\end{eqnarray}
	where $\varphi(2k)=\prod_{i=1}^{k}{w}_{-}(2i/A)/{w}_{+}(2i/A)$ and 
	$\varphi(0)\equiv 1$.  
	Since the termination time is dominated by $\tau(2)$, we can 
	use the  power laws for the annihilation and creation rates to derive an approximate closed 
	expression. For large values of $A$, the termination time 
	can be shown to be proportional to  \cite{vidmar2019extinction} 
	\begin{equation}
		\tau \sim {\rm exp} \left[ \frac{A}{2} \int_{2/A}^{n_{avg}} {\rm ln}  \frac{w_{+}(s)}{w_{-}(s)}
		ds \right]
		\label{scaling0}
	\end{equation} 
	which becomes, after substituting the expressions for $w_{\pm}$,
	\begin{equation}
		\tau \sim  \left( \frac{2}{A} \right ) ^{\nu_--\nu_+}
		e^{ A n_{avg} (\nu_--\nu_+)/2}\,,
		\label{scaling}
	\end{equation}
	Thus, consistent with earlier work using the cardiac models \cite{qu2006critical,vidmar2019extinction},
	the termination time increases exponentially with $A$.
	Importantly, aside from a prefactor that is independent of $A$, this expression provides an
	explicit estimate for
	$\tau$  for any domain size, including ones that would be prohibitively expensive to simulate directly. 
	In Fig.~\ref{fig:expdeptauonarea}C,  we plot the 
	values for $\tau$ obtained from the cardiac models as symbols together with a fit using Eq.~\ref{scaling0}.  The former ones were, 
	of course, only obtainable for small domain sizes while the fit can 
	be extended to arbitrarily large domain sizes.

	Finally, using the fitted power laws (Fig.~\ref{fig:statistics}B), we computed $\tau$ as a function of $a$ with the other parameters fixed to those corresponding to the cardiac models.
	We found that $\tau$ decreased by a factor of $\sim10^2$ in response to $a$ increasing by a mere factor of $10$, as shown in Fig.~\ref{fig:expdeptauonarea}D.  This great sensitivity of $\tau$ to $a$, considered together with the relative insensitivity of $a$ to $A$, suggests that changing $a$ is a mechanism for controlling the mean termination time for minimally varying domain sizes.
	Specifically, if modifying the electrophysiological parameters of the cardiac models increases attraction only, then $\tau$ decreases, resulting in shorter 
	termination times. 
	
\begin{figure}[h!]
		\begin{center}
	\includegraphics[width=\columnwidth]{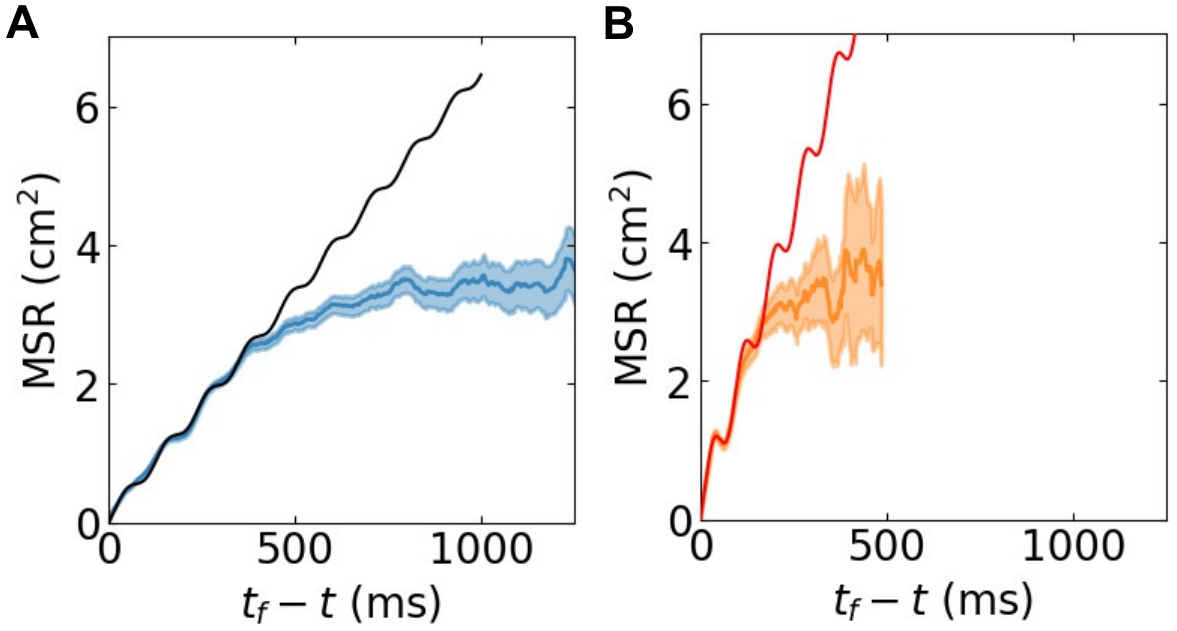}
			\caption{
				\textbf{A} MSR from the FK model with the excitability parameter, $\tau_d$, increased by 20\%.  The blue shaded region represents 95\% confidence intervals estimated via bootstrap.  The black line represents the fit to the OPM.
				\textbf{B} MSR from the LR model with the extracellular potassium concentration, $[K^+]_o$, decreased by 29.6\%.  The orange shaded region represents 95\% confidence intervals estimated via bootstrap.  The red line represents the fit to the OPM.
			}
			\label{fig:altering_attraction}
		\end{center}
	\end{figure}

	To verify that it is possible to vary $a$, and thus $\tau$, we  carried out additional simulations of the cardiac models. 
	For this, we altered 
	the excitability parameter $\tau_d$ in the FK model and the 
	extracellular potassium concentration $[K^+]_o$ in the LR model.
	We followed the same procedure as before and  fitted the MSR  expression for the OPM to the MSR curves obtained from the cardiac models 
	using simulated annealing on the last 300 ms (FK) and 100 ms (LR) before annihilation. 
	The fit results are shown in Figure~\ref{fig:altering_attraction} for both models. These fits, together with measured values of 
	$D$ and $\Gamma$, were then used to compute $a$ and the LPM model 
	was simulated to determine $\tau$.

	For the FK model, we increased $\tau_d$ by 20\%
	and obtained as
	fitted parameter values 
	$a_0=1.6077$ cm$^2/$s, 
	$a_1=1.3311$ cm$^2/$s, 
	$T_{OPM}=110.3665$ ms, and
	$\theta_f=-1.0730$ radians (RMSE=0.0459 cm$^2$).
	The measured parameters were $D=0.048\pm0.007$ cm$^2/$s and 
	$\Gamma=101.3\pm$1.3 ms, resulting in $a=1.82\pm0.01$ cm$^2/$s, which was 17.5\% larger than the corresponding value for $a$ in Table~\ref{tab:properties}. This led 
	to a decrease of the mean termination time by 35\%.
	In the LR model, we reduced $[K^+]_o$ 
	by 29.6\% and found the following
	fitted parameter values: 
	$a_0=4.1932$ cm$^2/$s, 
	$a_1=5.9593$ cm$^2/$s, 
	$T_{OPM}=82.4087$ ms,  and
	$\theta_f=-0.9490$ radians (RMSE=0.0352 cm$^2$). 
	Measured parameters were $D=0.42\pm0.03$ cm$^2/$s and 
	$\Gamma=45.9\pm1.1$ ms, resulting in $a=5.73\pm0.07$ cm$^2/$s, which was $38.6\%$ smaller than the corresponding value for $a$ in Table~\ref{tab:properties}. 
	The termination time in this case increased by $173.6\%$.
 Thus, 
 increasing $a$ reduced $\tau$ for the FK model, and
 reducing $a$ increased $\tau$ for the LR model.

	\section{Concluding Remarks}
	
	In summary, this study reveals that the annihilation dynamics of spiral waves
	in spatially extended 
	cardiac models 
	can be captured by a computationally efficient particle model.  In this model, the spiral wave tips are 
	represented by diffusing particles subject to a locally attractive force. 
	We showed that the particle model  
	accurately reproduced the annihilation rates of spiral tips in the cardiac models and that it can be used to efficiently model their tip dynamics.
	We also used the particle model to show that
	increasing the strength of the apparent attraction force accelerates annihilation, thus decreasing  
	the mean termination time.

 What have we ignored and how can this study be extended?
 First of all, we exclusively focused on annihilation dynamics. Clearly, 
 to construct more realistic particle models, we will have to include
 creation events. These events may require the inclusion of repulsion of spiral wave tips, 
	as has been reported
	in some cardiac models \cite{ermakova1989interaction}. 
 Once these events are incorporated into the model, we should be able
 to determine how parameters that describe these creation events can change
 termination times. 
 Second, we have simulated the dynamics on a domain 
 with periodic boundary conditions, guaranteeing that annihilation is a 
 pair-wise event. Introducing non-conducting boundaries will necessitate a description 
 of the annihilation of a single spiral wave that collides with one of these boundaries. 
 Third, we have simplified the oscillatory model to a linear model such that we did not have
 to compute the AAFs for the spiral waves. It would be interesting to extend our study
 to explicitly enforce conditions on the relative AAFs at the time of annihilation.  
	Finally, future work involving cardiac models can include 
	a more thorough investigation of  the dependence of the attraction coefficient between spiral tips on trans-membrane currents. 
	Reproducing desirable effects on these currents can then be the target of drug discovery,
	potentially opening a new door to noninvasive therapies for 
	clinically significant symptoms of cardiac fibrillation.
	
	\section{Acknowledgements}
	
	This work was supported by the National Institute of Health through awards 
	HL083359, HL122384, HL149134, and T32GM127235. T.T. acknowledges the Open Science Grid~\cite{osg07,osg09} for allocation of computational resources. P.M acknowledges the support of the Research Training in Mathematical and Computational Biology Grant (1148230) and the NSF Division of Mathematical Sciences award 1903275.
	
	\appendix
	\section{Detailed Equations and Parameters of the Full Cardiac Models}

	\subsection{Fenton-Karma Model}
	The ionic currents in  the Fenton-Karma (FK) cardiac model \cite{fenton1998vortex} are written as:  
	
	$$I_\text{ion}=I_\text{fi}+I_\text{so}+I_\text{si},$$
	
	\noindent where the fast inward current is
	
	$$I_\text{fi}(u,v)=-\frac{v}{\tau_d}(1-u)(u-u_c)\Theta(u-u_c),$$  
	
	\noindent the slow outward current is
	
	$$I_\text{so}(u)=u\frac{\Theta(u_c-u)}{\tau_0}+\frac{\Theta(u-u_c)}{\tau_r},$$
	
	\noindent and the slow inward current is
	
	$$I_\text{si}(u,w)=\frac{w}{2\tau_\text{si}}\Big(1+\tanh\big(k(u-u_c^\text{si})\big)\Big).$$ 
	
	\noindent 
	In these equations, $v$ and $w$ represent the fast and slow variable, respectively:
	\begin{equation}
		\frac{dv}{dt}=(1-v)\Theta(u_c-u)\bigg(\frac{\Theta(u-u_v)}{\tau_{v1}^-} + \frac{\Theta(u_v-u)}{\tau_{v2}^-}\bigg) -v\frac{\Theta(u-u_c)}{\tau_v^+}
	\end{equation}
	and 
	\begin{equation}
		\frac{dw}{dt}=(1-w)\frac{\Theta(u_c-u)}{\tau_w^-}-w\frac{\Theta(u-u_c)}{\tau_w^+}.
	\end{equation}
	Parameter values for the model were chosen to be 
	$\tau_d=0.45$ms, $\tau_0=12.5$ms, $\tau_r=$33.25ms, $\tau_\text{si}=29$ms, $k=10$,  $u_c^\text{si}=0.85$, $u_c=0.13$, $u_v=0.04$, $\tau_{v1}^-=1250$ms,
	$\tau_{v2}^-=19.6$ms,
	$\tau_v^+=13.03$ms,  $\tau_w^-=40$ms, and $\tau_w^+=800$ms.
	Using a diffusion coefficient of $D_u=0.0005$cm$^2$/ms resulted in   
	a conduction velocity of a planar wave of $c_v=51$cm/s, which is within the electrophysiological range.
	
	\subsection{Luo-Rudy Model}
	The total current for the Luo-Rudy (LR) cardiac model \cite{luo1991model} used in this 
	study is given by 
	
	$$I_\text{ion}=I_\text{Na}+I_\text{si}+I_\text{K1T}+I_\text{K},$$ 
	
	\noindent
	where the first term is the  sodium current
	
	$$
	I_\text{Na}=(u-E_\text{Na})G_\text{Na}jhm^3.
	$$
	
	\noindent with parameter values $G_\text{Na}=16\text{mS}/\text{cm}^2$ and $E_\text{Na}=54.4\text{mV}$.  
	The second term is the slow inward current 
	
	$$
	I_\text{si}=G_\text{si}(u-E_\text{si})fd,
	$$
	
	\noindent where $G_\text{si}=0.052\text{mS}/\text{cm}^2$ and $E_\text{si}=-82.3\text{mV}-13.0287\log\big([\text{Ca}^{+2}]_i)\big)$.
	The last two terms represent the 
	potassium current, which 
	is computed using the equations and parameters of Qu et al. in Ref.~\cite{qu2000local,qu2000mechanisms}.
	In these expression, 
	the intracellular calcium is given by 
	
	$$
	d[\text{Ca}^{+2}]_i/dt = -10^{-7} I_\text{si} + \big(0.07\big)\big(10^{-7}-[\text{Ca}^{+2}]_i\big)
	$$
	
	Furthermore, the dimensionless gating variables, $y=m,h,j,d,f$, 
	are described by equations of the form
	\begin{equation}\label{eqn:gating_var}
		dy/dt =(y_\infty-y)/\tau_y.
	\end{equation}
	where
	$$y_\infty(u)=a_y(u)\tau_y(u)$$
	and
	
	$$\tau_y(u)=\frac{1}{a_y(u)+b_y(u)},$$
	
	\noindent In these equations, $a_y$, $b_y$ are dimensionless  monotonic functions, 
	which were evaluated 
	in constant time using a lookup table based on the equations of Qu et al.
	\cite{qu2000local,qu2000mechanisms}.  
	Additional parameters can be found in these references.  For complete reproducibility, the equations and parameters that we used are included in Table~\ref{tab:lr_params} and here.  We have
 \begin{widetext}
	\begin{eqnarray*}
		I_\text{K1T}=I_{K1}(V_\text{m})+(1\mu \text{A}/\text{cm}^2)\frac{(V_\text{m}+87.95\text{mV})/54.6448\text{mV}}{1+e^{(7.488\text{mV}\, -V_\text{m})/5.98\text{mV}}} 
		+(1\mu \text{A}/\text{cm}^2)\frac{V_\text{m}+59.87\text{mV}}{25.5037\text{mV}},
	\end{eqnarray*}
 \end{widetext}
	where
	
 \begin{widetext}
 \begin{eqnarray*}
	I_{K1}(\Vm)=\frac{(1\mu\text{A}/\text{cm}^2)\frac{\Vm+87.95\mV}{1.62129\mV}{
			\sqrt{\frac{[K^+]_o}{5.4\text{mM}}}}}{
		\left(1+e^{\frac{V_\text{m}+28.735\mV}{4.19287\mV}}\right)
		\left(\frac{e^{\frac{\Vm-506.36\mV}{16.1943\mV}}+0.49124
			e^{\frac{V_\text{m}+93.426\text{mV}}{12.4502\mV}}}{1 +e^{-\frac{\Vm+92.703\mV}{1.94439\mV}}}+\frac{1.02\mV}{1
			+e^{\frac{V_\text{m}+28.735\mV}{4.19287\mV}}}\right)},
   \end{eqnarray*}
   \end{widetext}

	\noindent and where we have taken $[K^+]_o=5.4\text{mM}$.  
We computed $I_\text{K}$ using $I_\text{K}=I_1x$, where 
\begin{widetext}
 \begin{eqnarray*}
I_1=\Big(2.837G_\text{K}\sqrt{\frac{5.4\text{mM}}{[K^+]_o}}\Big)\frac{V_\text{m}-E_1}{\Vm+77\mV}\Big(\exp\Big(\frac{\Vm+77\mV}{25\mV}\Big)-1\Big)\Big/\exp\Big(\frac{\Vm+35\mV}{25\mV}\Big)\Big)
\end{eqnarray*}
   \end{widetext}
	
	\noindent where have used the values
	$G_\text{K}=0.423\text{mS}/\text{cm}^2$ and
	
	$$
E_{1}=(10^3\mV)\frac{RT}{F}\log \Big(
\frac{[K^+]_o+0.01833[Na^+]_o}{[K^+]_i+0.01833[Na^+]_i}\Big)\approx -77.61\mV.
	$$
	
	\noindent where we took $[K^+]_i=145\text{mM}$, $[{Na^+]_i=18\text{mM}}$, $[Na^+]_o=145\text{mM}$, $R = 8.3145   \text{J}/(\text{mol}\,^\circ\text{K})$ as the universal gas constant, and $F=96485.3321233100184 \text{C}/\text{mol} $ as Faraday's constant.
	We supposed a homeostatic body temperature of $T=37^\circ C$ fixed.  We have similarly approximated 
	
	$$
E_{\text{K}1}=(10^3\mV)\frac{RT}{F}\log\bigg(\frac{[K^+]_o}{[K^+]_i}\bigg)\approx 87.95\mV.
	$$
	
To avoid numerical overflow when time evolving these gating variables, we treated time scales as zero, $\tau_y=0$, whenever they took a value $\tau_y\le 5\cdot 10^{-4}$ms.  Additionally, we evaluated $I_1$ using L'Hospital's Rule in the $10^{-6}\mV$ neighborhood of $\Vm=-77\mV$.

	Finally, for 
	the diffusion coefficient we used $D_u=0.001$cm$^2$/ms, 
	resulting in a 
	conduction velocity of $c_v=33$cm/s, which is within the electrophysiological range.

\begin{table*}
\caption{Parameters of the Luo-Rudy-I model used in our study. For ease of notation, we used the dimensionless transmembrane voltage field  $u=V_m/(1mV)$. Gating variables time evolve according to Eq. S\ref{eqn:gating_var}.}
\label{tab:lr_params}
\begin{ruledtabular}
	\begin{tabular}{@{}ccc@{}}
 \hline
 \multicolumn{1}{c}{$y$} & \multicolumn{1}{c}{$a_y$} & \multicolumn{1}{c}{$b_y$} \\ 
 \hline
 m     &        $\frac{u+47.13}{3.125}\Big/\Big(1-e^{-\frac{u+47.13}{10}}\Big)$
&   $0.08 e^{-u/11}$                            \\
h     &   $0.135\exp\Big(\frac{u+80}{6.8}\Big)\Theta(40-u)$         
		&      \makecell{$\frac{100}{13}\frac{\Theta(40-u)}{1+\exp\big(\frac{10.66-u}{11.1}\big)}\; $\\ $+\;\Big(3.56e^{0.079u}+3.1\cdot 10^5 e^{0.35u}\Big)
		\Theta(u-40)$     } \\ \midrule%
	j       &    $-(u+37.78)\frac{\left(3.474\cdot 10^{-5}
	e^{-0.04391u}+127140
	e^{0.24444u}\right)}{1 +e^{0.311
	(u+79.23)}}\Theta(40-u)$        
&    \makecell{$\frac{0.3
	e^{-2.535\cdot 10^{-7}
		u}}{1+e^{-0.1 (u+32)}}\Theta(u-40) $ \\+
$ 0.1212 e^{-0.01052u}\frac{\Theta(40-u)}{1+e^{-0.1378(u+40.14)}}$}  \\ \midrule%
d      &   $0.095 e^{0.01(5-u)}\Big/\Big(1+e^{0.072(5-u)}\Big)$         
			&      $0.07 e^{-0.017(44+u)}\Big/\Big(1+e^{0.05(44+u)}\Big)$      \\ \midrule%
f      &     $0.012 e^{-0.008
(28+u)}\Big/\Big(1+e^{0.15(28+u)}\Big)$       &    $0.0065 e^{-0.02(30+u)}\Big/\Big(1.\,+e^{-0.2(30+u)}\Big)$        \\ \midrule%
x     &          $0.5\cdot 10^{-3}e^{0.083(u+50)}
			\Big/\Big(1+e^{0.057(u+50)}\Big)$               
			&           $1.3\cdot 10^{-3}\exp\big(\frac{u+20}{25}\big)
\Big/\Big(1+\exp\big(\frac{3}{2}\frac{u+20}{25}\big)\Big)$                     \\ 
 \end{tabular}
 \end{ruledtabular}
 \end{table*}

\end{document}